\documentclass[twocolumn,amsmath,amssymb,prl,showkeywords]{revtex4}
\usepackage{mathrsfs}
\usepackage{graphicx,color}
\usepackage{dcolumn}
\usepackage{bm}
\usepackage{CJK}
\usepackage{epstopdf}
\usepackage{changes}

\begin{document}
\begin{CJK*}{GBK}{song}
\title{Octupole deformation in even-even Ra isotopes from covariant density functional theory with localized exchange terms in a three-dimensional lattice space}

\author{Z. Y. Dong$^{1}$}
\author{Z. X. Ren$^{2}$}
\author{Q. Zhao$^{3}$}\email{zhaoqiang@ciae.ac.cn}
\author{Z. M. Niu$^{1}$}\email{zmniu@ahu.edu.cn}

\affiliation{$^1$School of Physics, Anhui University,
             Hefei 230601, China}
\affiliation{$^2$School of Physics, Nankai University,
             Tianjin 300071, China}
\affiliation{$^3$China Institute of Atomic Energy,
             Beijing 102413, China}

\begin{abstract}
The covariant density functional theory in a three-dimensional lattice space is extended to the PCF-PK1 functional with localized exchange terms and is employed to study the nuclear shape evolution of even-even Ra isotopes. Well-developed axial octupole deformations are found for the ground states of $^{222-228}$Ra with no evidence of triaxial shapes. The energy gain of octupole deformation is employed to assess the stability of octupole deformation, with relatively larger values observed for $^{224}$Ra and $^{226}$Ra. A simplified analysis method based on the single-particle spectrum at the octupole deformation parameter $\beta_3=0$ is proposed to identify the key single-particle levels driving octupole deformation. It is found that the $m_z=3/2$ orbitals from $\nu 1j_{15/2}$ and $\nu 2g_{9/2}$ and the $m_z=1/2$ orbitals from $\pi 1i_{13/2}$ and $\pi 2f_{7/2}$, play crucial roles in the formation of octupole deformation in Ra isotopes. Furthermore, increasing the tensor coupling strength promotes octupole deformation, whereas reducing the pairing strength stabilizes it. Our results provide a microscopic understanding of octupole deformation in the Ra isotopic chain and highlight the importance of both tensor coupling and pairing correlations in reflection-asymmetric nuclear shapes.
\end{abstract}


\maketitle

\section{Introduction}
The atomic nucleus, as a finite quantum many-body system, exhibits a variety of shapes due to spontaneous symmetry breaking. These shapes can be described by a multipole expansion of the nuclear surface in terms of spherical harmonics~\cite{Bohr1975Book}. Although the quadrupole deformation is the dominant mode in nuclei, higher multipole deformations play an essential role in many nuclear phenomena, such as fission barriers~\cite{Zhou2016PS, Lu2012PRC, Rutz1995NPA}, parity-doublet bands in excitation spectra~\cite{Chen2008PRC, Butler1996RMP}, and high-$K$ isomers~\cite{Liu2013PRC, Liu2011PRC}. The reflection-asymmetric octupole deformation manifesting as a pear shape has recently attracted significant attention, since it not only significantly influences nuclear structure, fission dynamics, and the occurrence of parity-doublet bands, but also provides a unique platform to search for  the Charge-Parity violation via atomic electric dipole moments~\cite{Butler1996RMP, Chupp2019RMP, Engel2025ARNPS}.

From a microscopic perspective, the octupole deformation arises from the coupling of single-particle orbitals with opposite parity and satisfying $\Delta l = \Delta j = 3$ near the Fermi surface~\cite{Nazarewicz1984NPA, Moller2008ADNDT}. This condition is typically satisfied when the proton number is around 34, 56, or 88, and the neutron number is around 34, 56, 88, or 134. For example, in the nuclei with $Z\approx 88$ and $N\approx 134$, the neutron orbital pair ($\nu g_{9/2}, \nu j_{15/2}$) and the proton orbital pair ($\pi f_{7/2}, \pi i_{13/2}$) are located close to the Fermi surface, generating strong octupole correlations and potentially  leading to a pear-like shape~\cite{Nazarewicz1984NPA, Moller2008ADNDT}.

Experimentally, the reflection asymmetry in nuclei can be probed by measuring the collective electric octupole transition strength and the enhanced electric dipole transitions between opposite-parity doublet bands~\cite{Butler1996RMP,Cottle1990PRC,Spear1990PRC,Butler2019NatCom}. Static octupole deformation has been experimentally verified in some nuclei, including $^{144,146}$Ba~\cite{Bucher2016PRL, Bucher2017PRL} and $^{222,224,226}$Ra~\cite{Butler2020PRL, Gaffney2013Nature, Wollersheim1993NPA}. In contrast, $^{228}$Ra behaves consistently with the octupole vibrational model, exhibiting strong octupole correlations~\cite{Butler2020PRL}. Extensive experimental efforts have made the Ra isotopic chain a hot topic of continuous interest in the octupole deformation studies.

Theoretically, the octupole deformation has been studied using various approaches, including the finite-range droplet model (FRDM)~\cite{Moller2008ADNDT}, covariant density functional theory (CDFT)~\cite{Agbemava2016PRC, Agbemava2017PRC, Zhao2024PRC, Zhang2019PRC}, Hartree-Fock-Bogoliubov theory with Gogny interactions~\cite{Robledo2011PRC, Robledo2012JPG, Robledo2015JPG} and Skyrme interactions~\cite{Erler2012PRC, Cao2020PRC, Loc2023PRC, Ebata2017PS}, cranked shell model~\cite{Nazarewicz1985NPA, He2020PRC}, projected shell model~\cite{Chen2000PRC, Chen2015PRC}, cluster model~\cite{Shneidman2003PRC, Buck2008JPG, Shneidman2015PRC}, interacting boson model~\cite{Nomura2013PRC, Nomura2014PRC}, and generator coordinate method~\cite{Li2013PLB, Rong2023PLB, Yao2015PRC, Yao2016PRC}. The CDFT, in particular, is distinguished by its Lorentz covariance, which naturally generates the spin-orbit interaction without adjustable parameters, consistently treats time-odd fields, and provides a natural explanation for the pseudo-spin symmetry observed in nuclei~\cite{Niksic2011PPNP, Vretenar2005PR, Ring1996PPNP, Bender2003RMP}. These features make the CDFT a well-suited tool for describing both ground-state and excited-state properties of nuclei. However, most previous CDFT calculations have been performed under some restricted space symmetry assumptions, which may affect the discovery of novel nuclear shapes~\cite{Zhou2016PS, Lu2014PRC, Zhao2012PRC, Zhao2017PRC}. To overcome this limitation, a three-dimensional (3D) lattice approach has been developed within CDFT, which enables the description of arbitrary nuclear shapes without any space symmetry restrictions~\cite{Ren2017PRC, Ren2019SCPMA, Hagino2010PRC, Tanimura2015PTEP}. This method has been successfully applied to study nuclear exotic shapes such as the linear-chain structures in $^{12}\mathrm{C}$~\cite{Ren2019SCPMA}, as well as tetrahedral shapes in $^{110}$Zr and $^{286}$No~\cite{Xu2024PRC, Xu2024PLB}.

Besides the spatial symmetry restrictions, most conventional CDFT calculations neglect the exchange (Fock) terms and lack tensor couplings. These can be overcome by explicitly incorporating the exchange terms~\cite{Zhao2024PLB, Zhao2023PLB, Zhao2022PRC}. In particular, the recently developed point-coupling density functional PCF-PK1 incorporates localized exchange terms via the Fierz transformation and introduces tensor interactions in the Lagrangian, while preserving the computational simplicity of point-coupling models~\cite{Zhao2022PRC}. Consequently, PCF-PK1 brings several key improvements. First, it improves the description of single-particle levels by eliminating the spurious shell closures at $Z = 58$ and $92$, which exist commonly in many covariant density functionals without exchange terms. Second, the PCF-PK1 gives reasonable spin-orbit splittings in finite nuclei with larger Dirac mass and Landau mass due to the inclusion of tensor couplings, which is associated with the higher level densities around the Fermi surface. In addition, it reproduces Gamow-Teller resonances without any adjustable parameters, demonstrating that a self-consistent description of spin-isospin excitations can be achieved with the localized exchange terms~\cite{Zhao2022PRC}.

Since the octupole deformation involves shape degrees of freedom that break reflection symmetry, a 3D description without spatial symmetry restrictions is essential to study complicated nuclear shapes. The recently developed PCF-PK1 functional, which incorporates localized exchange terms and tensor couplings, can reliably describe nuclear single-particle spectra essential to the study of octupole deformation. Therefore, the CDFT with localized exchange terms in a 3D lattice space offers an ideal framework to systematically investigate octupole deformation~\cite{Butler1996RMP}. In this work, the CDFT in a 3D lattice space is extended to the PCF-PK1 functional with localized exchange terms and is employed to study the nuclear shape evolution of even-even Ra isotopes in the mass range $A = 218$ to $240$. We calculate the potential energy surfaces, analyze the microscopic mechanism of octupole deformation, and special attention will be focused on the effects of the tensor coupling strength and pairing correlations on octupole deformation. The corresponding results are given in Sec.~III. The framework is presented in Sec.~II. Finally, the summary and perspectives are given in Sec.~IV.

\section{Theoretical framework}
The starting point of the CDFT is an effective Lagrangian density, and we employ the point-coupling Lagrangian density with PCF-PK1 effective interaction~\cite{Zhao2022PRC} in this work. Besides the $S$, $V$, $tS$, and $tV$ channels already included in the conventional point-coupling model~\cite{Nikolaus1992PRC, Burvenich2002PRC, Zhao2010PRC, Liu2023PLB}, the PCF-PK1 functional additionally incorporates the $T$, $tT$, $PS$, $tPS$, $PV$, and $tPV$ channels, where $S$, $V$, $T$, $PS$, and $PV$ denote the scalar, vector, tensor, pseudoscalar, and pseudovector couplings, respectively, and the subscript ``$t$'' refers to the corresponding isovector channel.

The corresponding Hamiltonian can be obtained with the Legendre transformation. Based on the no-sea approximation and the Hartree-Fock approximation, the energy density functional can be calculated as the expectation value of the Hamiltonian in the ground-state Slater determinant~\cite{Zhao2010PRC, Zhao2022PRC}. Using the Fierz transformation, the exchange terms can be simply expressed as the superposition of the Hartree terms. According to the conventional variational principle on the energy density functional, one obtains the Dirac equation for nucleons,
\begin{equation}
[\boldsymbol{\alpha} \cdot \boldsymbol{p} + \beta(M + S) + V^0 - \boldsymbol{\alpha} \cdot \boldsymbol{V} - i\beta\boldsymbol{\alpha} \cdot \boldsymbol{T}^0 ] \psi_k
= \varepsilon_k \psi_k,
\end{equation}
where $\varepsilon_k$ is the single-particle energy, $M$ is the nucleon mass. The scalar potential $S$, vector potential $V^\mu$, and tensor potential $\boldsymbol{T}^0$ are connected in a self-consistent way to the corresponding nucleon densities and currents~\cite{Zhao2022PRC}.

\begin{figure}[h]
\includegraphics[width=7.6cm]{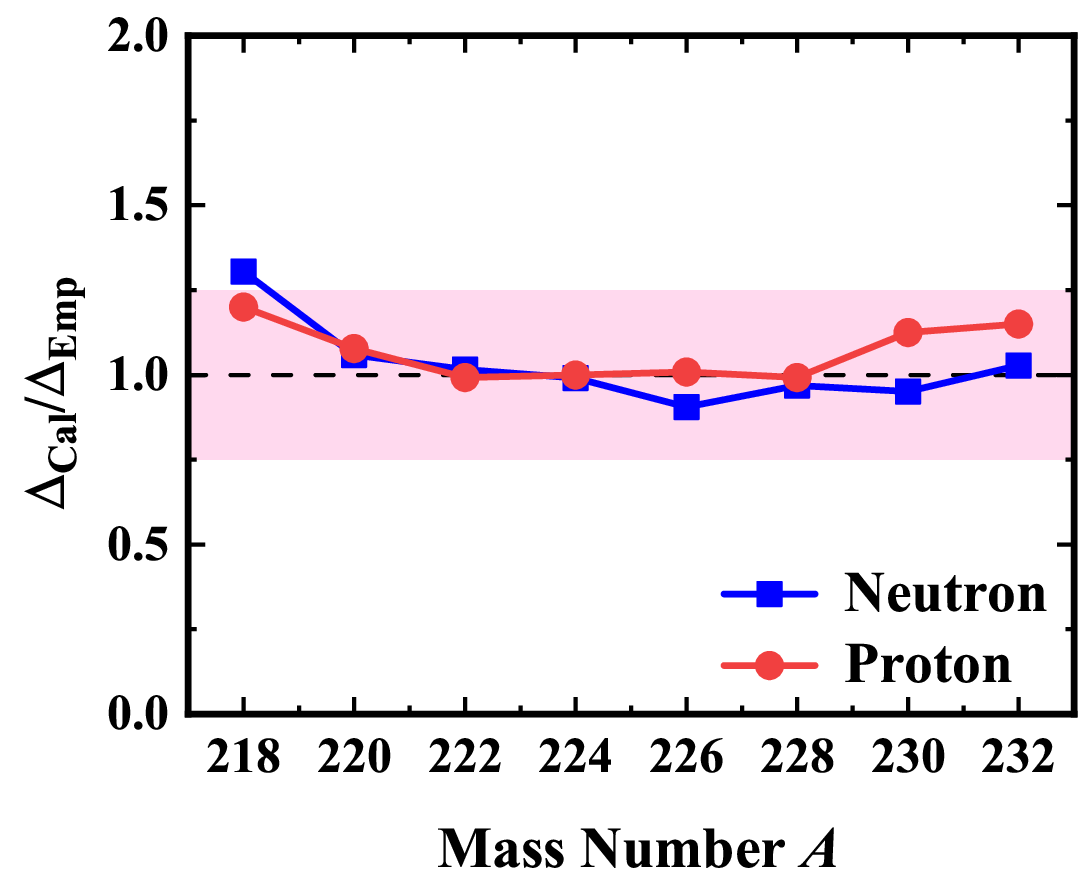}
\caption{(Color online) Ratios of the pairing gaps calculated by the covariant density functional theory in 3D lattice space to the empirical values extracted from the three-point odd-even mass differences for even-even $^{218-232}$Ra nuclei~\cite{Wang2013PRC}. The experimental masses are taken from the AME2020 atomic mass evaluation~\cite{Wang2021CPC}.}
\label{Fig:gapratio}
\end{figure}

Pairing correlations are treated within the Bardeen-Cooper-Schrieffer (BCS) method~\cite{RingSchuck1980Book}. The pairing energy functional is given by
\begin{eqnarray}
E_{\mathrm{pair}} = \sum_{\tau=n,p} \frac{G_{\tau}}{4} \int d^3r \, \kappa_{\tau}^*(\mathbf{r}) \kappa_{\tau}(\mathbf{r}),
\end{eqnarray}
where pairing strength parameter $G_{\tau}$ is negative, indicating the attractive nature of the pairing interaction, and $\kappa(\mathbf{r})$ is the pairing tensor~\cite{Xu2024PLB}. In this work, the pairing strengths for neutrons and protons are taken as $G_n = -277$~MeV$\cdot$fm$^{3}$ and $G_p = -340$ MeV$\cdot$fm$^{3}$, respectively, which are determined by optimizing the ratio of the calculated pairing gaps to the empirical values as close to unity as possible for even-even $^{218-232}$Ra nuclei (see Fig.~\ref{Fig:gapratio}).

The Dirac equation is solved on a 3D lattice space. The step sizes and the grid numbers along the $x$, $y$, and $z$ axes are taken as $0.8$ fm and $34$, respectively. To overcome the variational collapse and fermion doubling problems, the inverse Hamiltonian method~\cite{Hagino2010PRC} and the Fourier spectral method~\cite{Tanimura2015PTEP} are employed for solving the Dirac equation~\cite{Ren2017PRC}. In contrast to previous relativistic calculations, no space symmetry restrictions are imposed on our solutions, allowing arbitrary nuclear shapes to be described self-consistently~\cite{Ren2019SCPMA,Xu2024PRC}.

For axially symmetric quadrupole and octupole deformations, the intrinsic shape is characterized by the deformation parameters $\beta_{2}$ and $\beta_{3}$, which can be determined from the following relations
\begin{eqnarray}
\beta_{2} = \frac{4\pi}{3 A R^2} \int d^3r \, \rho_V(\mathbf{r}) r^2 Y_{20}(\Omega), \\
\beta_{3} = \frac{4\pi}{3 A R^3} \int d^3r \, \rho_V(\mathbf{r}) r^3 Y_{30}(\Omega),
\end{eqnarray}
where $R = 1.2 A^{1/3}$ fm, $A$ is the mass number, and $\rho_V(\mathbf{r})$ is the vector density~\cite{Zhou2016PS}. These parameters describe the quadrupole-deformed (prolate/oblate) and octupole-deformed (pear-like) shapes of the nucleus, respectively.

\section{Results and discussion}
\begin{figure}[htbp]
\includegraphics[width=0.49\textwidth]{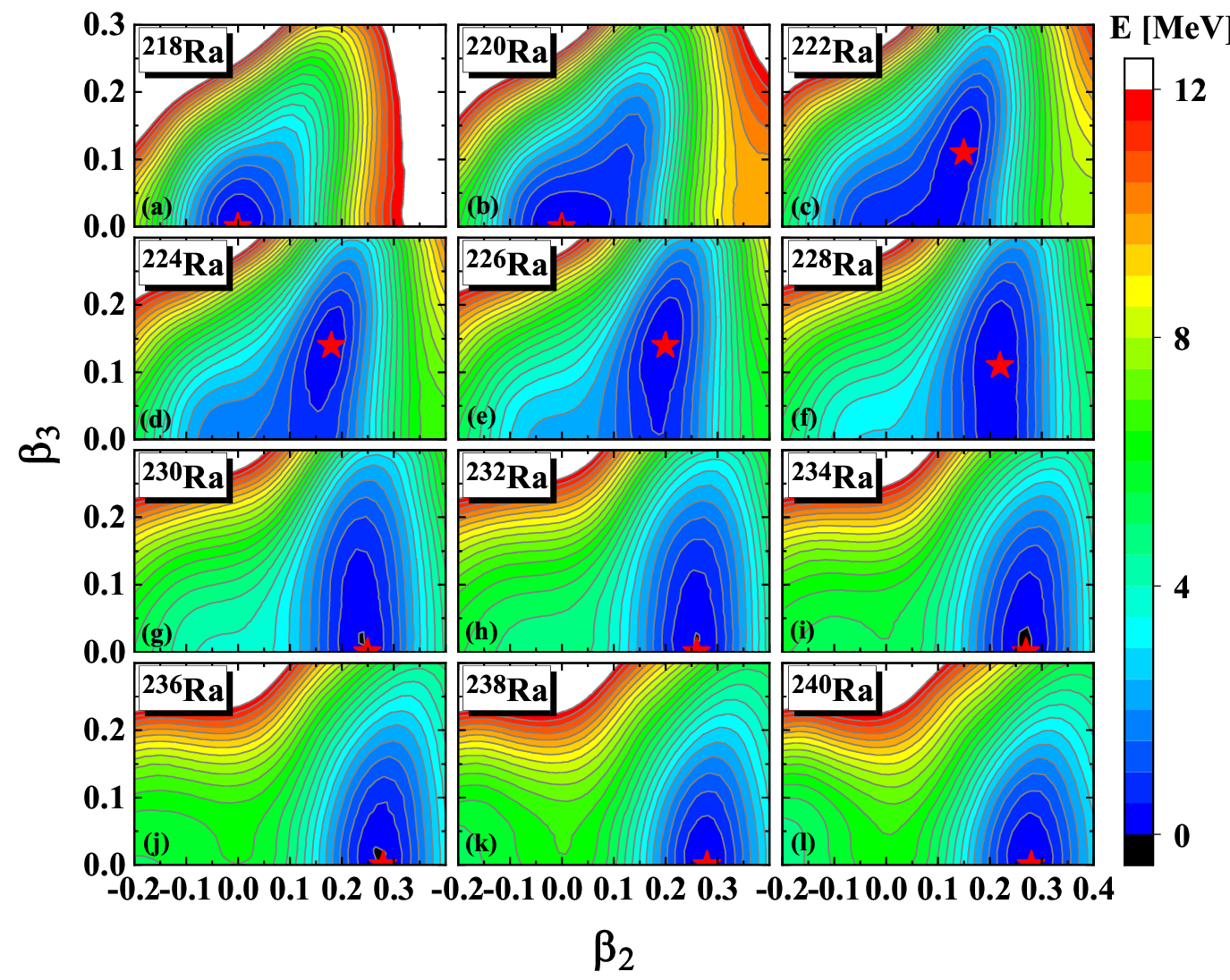}
\caption{(Color online) The potential energy surfaces of Ra isotopes in the $(\beta_2, \beta_3)$ plane. The energies are normalized with respect to the ground-state energy (indicated by a red star), and the contour interval is 0.5~MeV.}
\label{Fig:PES}
\end{figure}
The potential energy surfaces (PESs) of Ra isotopes in the $(\beta_2, \beta_3)$ plane are shown in Fig.~\ref{Fig:PES}. It can be seen that $^{218}$Ra and $^{220}$Ra exhibit spherical ground states. Octupole deformation begins to develop in the Ra isotopes from $^{222}$Ra, although the octupole minimum remains soft. For $^{224}$Ra and $^{226}$Ra, well-developed octupole deformations emerge, characterized by a pronounced octupole minimum and a stiff potential in the octupole direction. As the neutron number further increases, the octupole softness reappears in $^{228}$Ra. Based on our calculations, $^{222-228}$Ra exhibit axial octupole deformation without triaxiality. From $^{230}$Ra onwards, the nuclei evolve into stable prolate quadrupole shapes.

\begin{figure}[htbp]
\includegraphics[width=7.6cm]{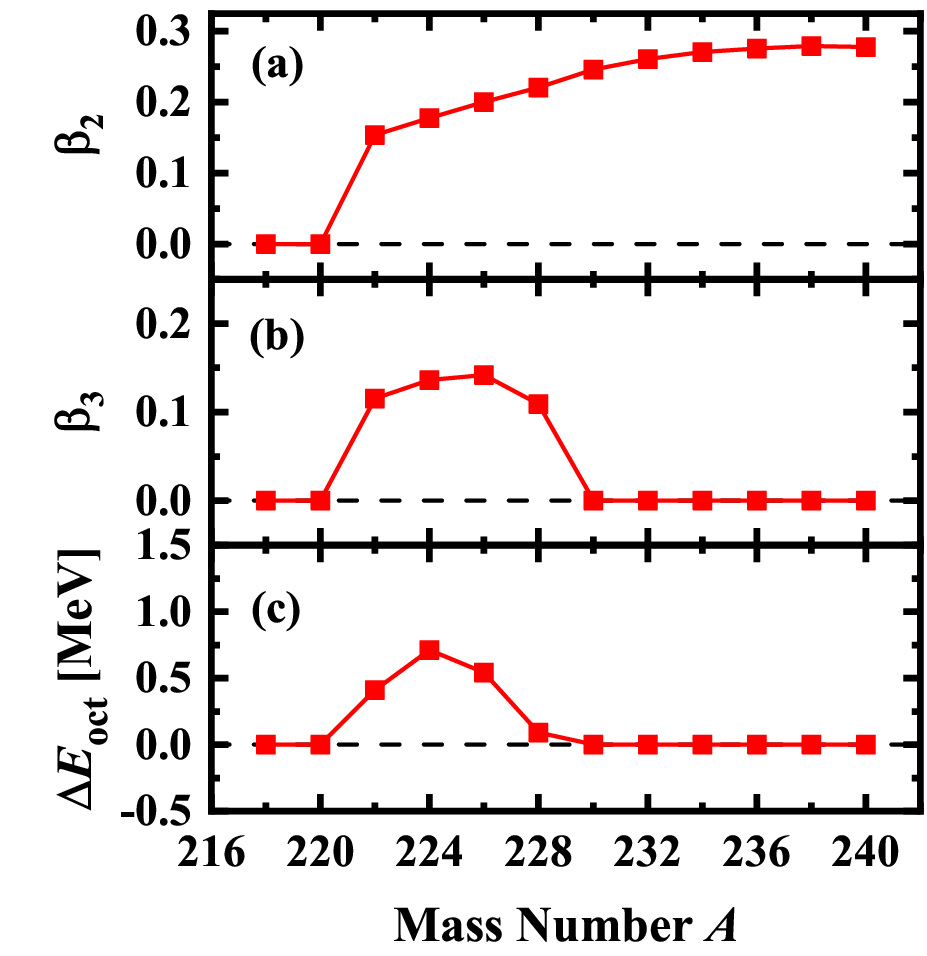}
\caption{(Color online) The calculated ground-state quadrupole $\beta_2$ (top panel), octupole $\beta_3$ (middle panel) deformations, and the energy gain of octupole deformation energy $\Delta E_{\mathrm{oct}}$ (bottom panel) are presented as functions of the mass number $A$. The results are obtained using the PCF-PK1 point-coupling density functional (red solid line).}
\label{Fig:betaEoct}
\end{figure}
Figure.~\ref{Fig:betaEoct} displays the calculated $\beta_2$, $\beta_3$, and $\Delta E_{\mathrm{oct}}$ as functions of the mass number $A$. Here
\begin{eqnarray}
\Delta E_{\mathrm{oct}} = E_{\mathrm{oct}} - E_{\mathrm{quad}}
\end{eqnarray}
quantitatively characterizes the effect of octupole deformation, where $E_{\mathrm{oct}}$ and $E_{\mathrm{quad}}$ are the binding energies at the octupole-deformed minimum and the minimum with no octupole deformation, respectively. Large $\Delta E_{\mathrm{oct}}$ value are typical for well-pronounced octupole minima in the PES, indicating a static octupole deformation, while small $\Delta E_{\mathrm{oct}}$ value are characteristic of octupole-soft PESs typical of octupole vibrations. Clearly, the quadrupole deformation $\beta_2$ increases gradually from $^{222}\mathrm{Ra}$ onward, reflecting a transition from spherical to well-deformed shapes. Both $\beta_3$ and $\Delta E_{\mathrm{oct}}$ are nonzero only within the narrow mass region $^{222-228}\mathrm{Ra}$ and vanish rapidly outside this interval. Among these isotopes, $^{222-226}\mathrm{Ra}$ exhibit relatively large $\Delta E_{\mathrm{oct}}$ values, indicating well-pronounced octupole minima and static octupole deformation. It is noteworthy that although $^{222}\mathrm{Ra}$ and $^{228}\mathrm{Ra}$ possess similar $\beta_3$ values, the $\Delta E_{\mathrm{oct}}$ value of $^{228}\mathrm{Ra}$ is considerably smaller, reflecting a much softer PES in the octupole direction, typical of octupole vibrations.

\begin{figure}[htbp]
\includegraphics[width=0.49\textwidth]{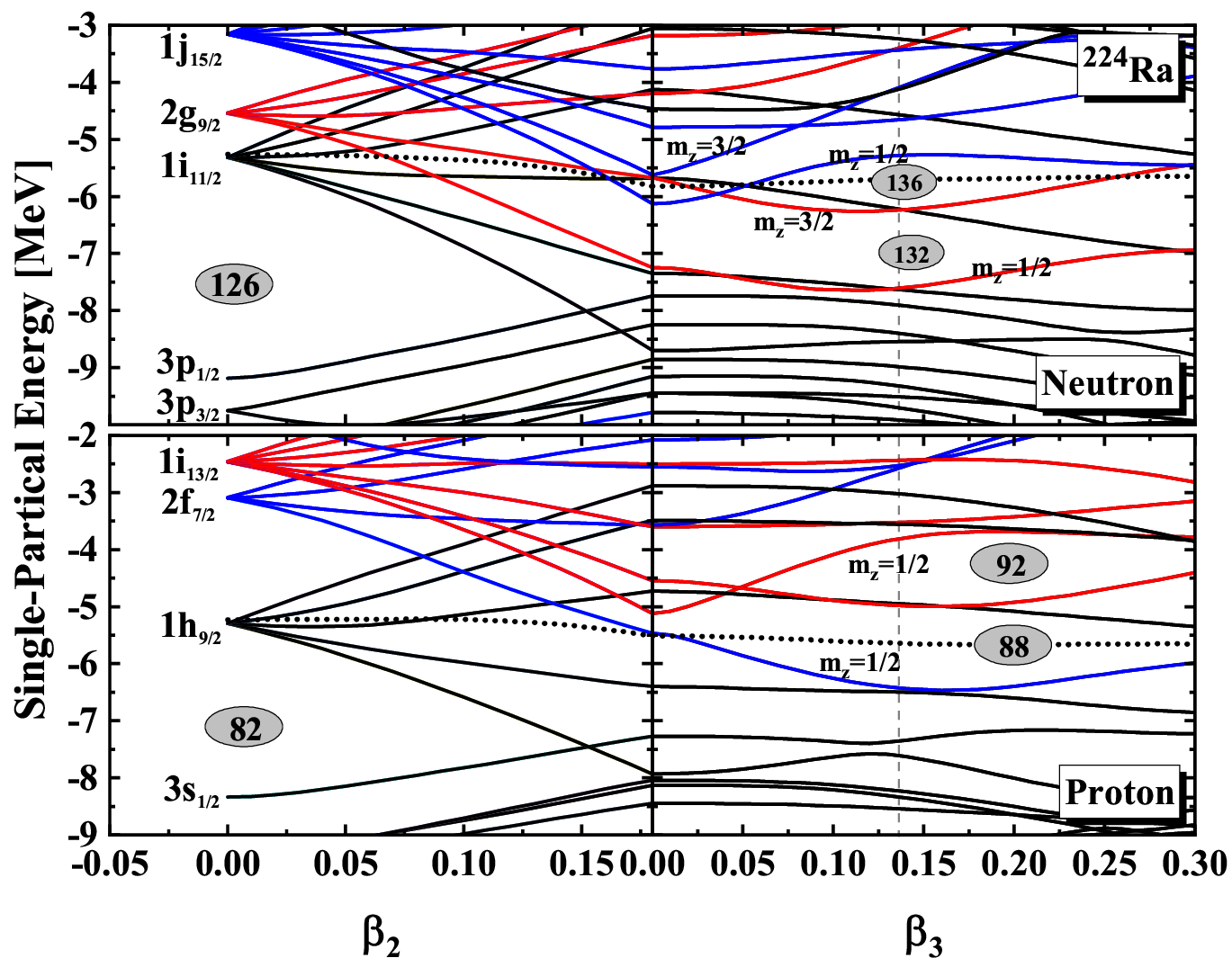}
\caption{(Color online) Single-neutron levels (top panel) and single-proton levels (bottom panel) of $^{224}$Ra as functions of the deformation parameters, calculated with the PCF-PK1 functional. The left panels illustrate the evolution with quadrupole deformation $\beta_2$ up to the ground-state value of $0.18$ with octupole deformation fixed at $\beta_3 = 0$. The right panels display the variation of single-particle energies as a function of octupole deformation $\beta_3$ from $0$ to $0.3$ with quadrupole deformation fixed at $\beta_2 = 0.18$. The Fermi levels are indicated by black short-dotted curves.}
\label{Fig:SPE}
\end{figure}
For a microscopic understanding of octupole deformation, we analyze the single-particle level structure of a representative nucleus $^{224}\mathrm{Ra}$. Figure~\ref{Fig:SPE} shows the single-neutron and single-proton levels as functions of the deformation parameters. The left panels illustrate the evolution with quadrupole deformation $\beta_2$ up to the ground-state value of $0.18$ (obtained from the global minimum of the PES) with octupole deformation fixed at $\beta_3 = 0$. The right panels display the dependence of single-particle energies on octupole deformation $\beta_3$ from 0 to 0.3 with quadrupole deformation fixed at $\beta_2 = 0.18$. The formation of the octupole minimum in $^{224}\mathrm{Ra}$ can be attributed to the coupling between the $m_z = 3/2$ neutron orbitals from $\nu 1j_{15/2}$ and $\nu 2g_{9/2}$ near the Fermi surface, while the $m_z = 1/2$ neutron orbitals also play a weaker role, as well as the coupling between the $m_z = 1/2$ proton orbitals from $\pi 1i_{13/2}$ and $\pi 2f_{7/2}$. At $\beta_2 = 0.18$, as $\beta_3$ increases from 0 to about 0.14, this octupole coupling induces pronounced level repulsion between these relevant orbitals near the Fermi surface. This reduces the local single-particle level density and hence provides an additional binding energy for the octupole-deformed minimum in $^{224}\mathrm{Ra}$, thereby stabilizing the octupole deformation.

\begin{figure}[htbp]
\includegraphics[width=8.6cm]{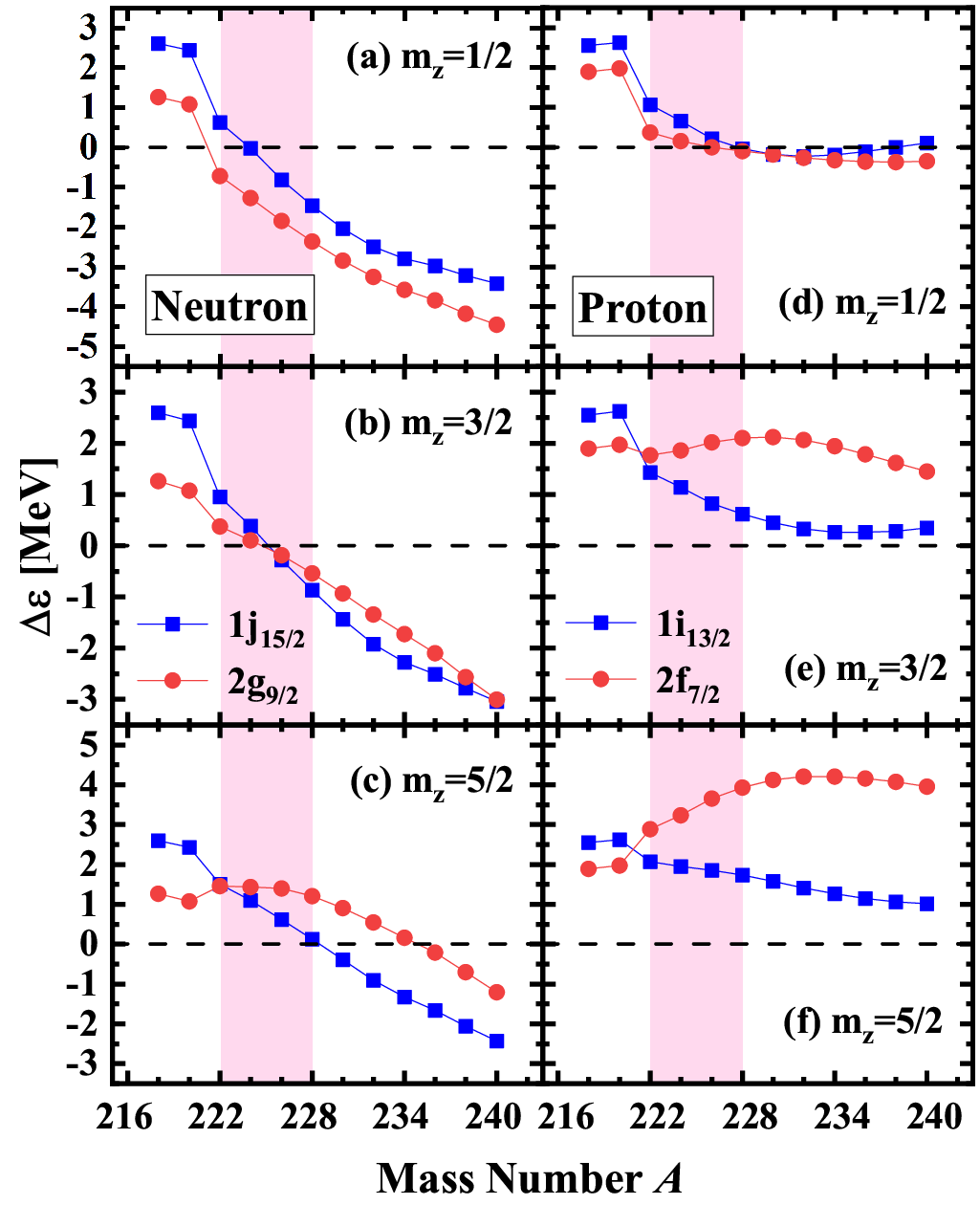}
\caption{(Color online) Single-particle energies of neutron orbitals from $\nu 1j_{15/2}$ and $\nu 2g_{9/2}$, and proton orbitals from $\pi 1i_{13/2}$ and $\pi 2f_{7/2}$, with $m_z = 1/2$, $3/2$, and $5/2$, near the Fermi surface for even-even Ra isotopes, relative to the Fermi level. The results are obtained from CDFT calculations in 3D lattice space using the PCF-PK1 functional at the quadrupole deformation $\beta_2$ optimized to the energy minimum and the octupole deformation constrained to $\beta_3 = 0$. The pink shaded band indicates the octupole-deformed nuclei identified in the previous calculations.}
\label{Fig:nobeta3min}
\end{figure}
The single-particle levels at the ground-state quadrupole deformation $\beta_2$ with $\beta_3 = 0$ play crucial roles in studying the microscopic mechanism of octupole deformation. Figure~\ref{Fig:nobeta3min} shows the single-particle energies of neutron orbitals from $\nu 1j_{15/2}$ and $\nu 2g_{9/2}$, and proton orbitals from $\pi 1i_{13/2}$ and $\pi 2f_{7/2}$, with $m_z = 1/2$, $3/2$, and $5/2$, near the Fermi surface for $^{218-240}\mathrm{Ra}$, relative to the Fermi level. If there are $\Delta l = \Delta j = 3$ orbital pairs with a small energy difference near the Fermi surface for both neutrons and protons, the strong coupling between these orbitals could induce nuclear octupole deformation. For $^{222-228}\mathrm{Ra}$ ($N = 134-140$), the $m_z = 3/2$ neutron orbitals from $\nu 1j_{15/2}$ and $\nu 2g_{9/2}$, together with the $m_z = 1/2$ proton orbitals from $\pi 1i_{13/2}$ and $\pi 2f_{7/2}$, are located close to the Fermi surface with a small energy difference, thereby providing favorable conditions for octupole coupling. In contrast, for $^{218,220}\mathrm{Ra}$ and $^{230-240}\mathrm{Ra}$, these orbitals are either located far from the Fermi surface or exhibit larger energy differences, which suppresses the formation of octupole deformation. Notably, although the $m_z = 1/2$ proton orbitals satisfy the proximity and small energy difference conditions for $^{230-240}\mathrm{Ra}$, the absence of suitable neutron orbitals with similar properties prevents the development of octupole deformation. The above analyses help us to understand the microscopic mechanism of octupole deformation by combining the potential energy surfaces in Fig.~\ref{Fig:PES}.

\begin{figure}[htbp]
\includegraphics[width=8.6cm]{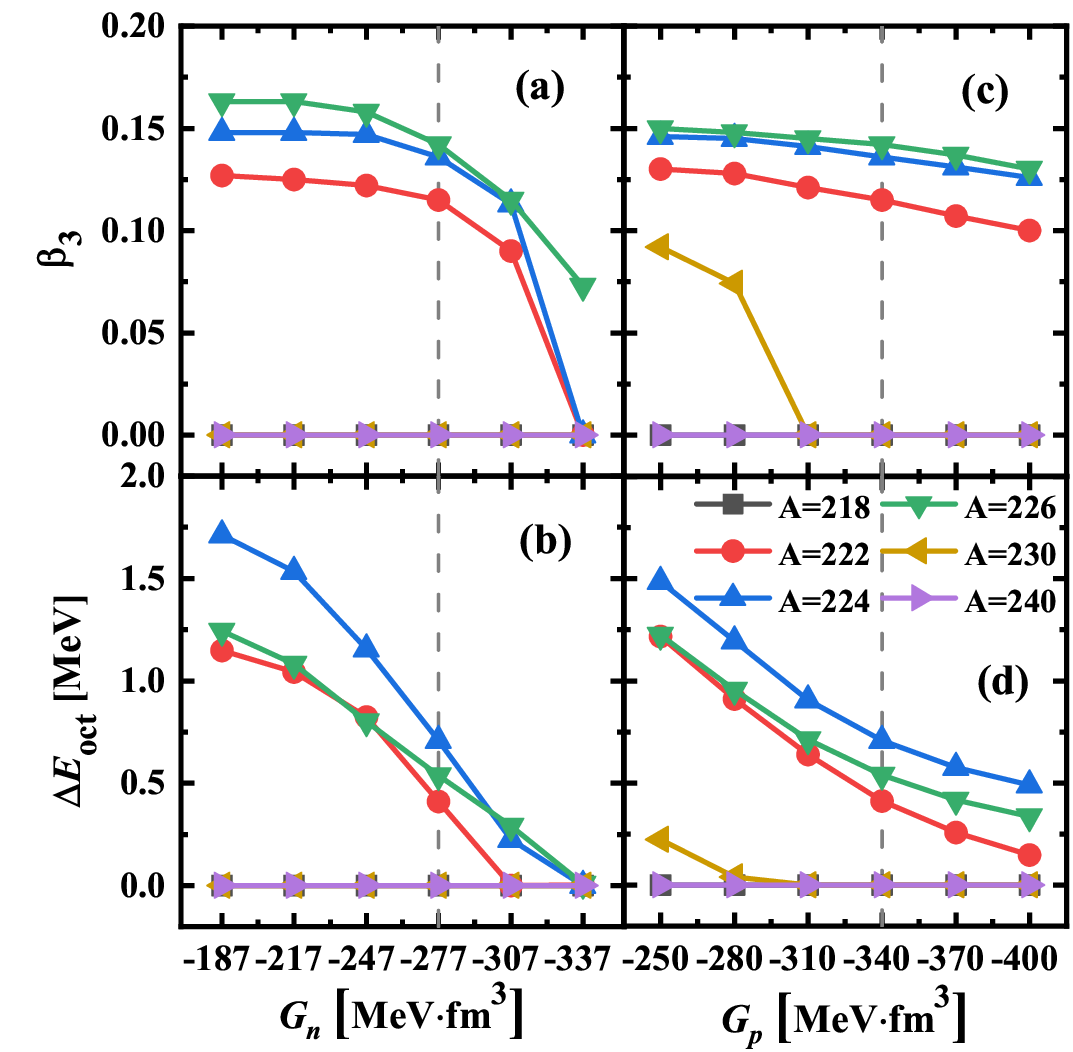}
\caption{(Color online) The calculated ground-state octupole deformation $\beta_3$ and the energy gain of octupole defomation $\Delta E_{\mathrm{oct}}$ as functions of the neutron (proton) pairing strength parameters $G_n$ ($G_p$) for Ra isotopes.}
\label{Fig:Vpair}
\end{figure}
The effect of pairing correlations on the octupole deformation is illustrated in Fig.~\ref{Fig:Vpair}, which presents the variations of ground-state octupole deformation $\beta_3$ and the energy gain of octupole defomation $\Delta E_{\mathrm{oct}}$ with the neutron and proton pairing strength parameters $G_n$ and $G_p$. A reduction in pairing strength leads to an increase in both $\beta_3$ and $\Delta E_{\mathrm{oct}}$, indicating a more stable octupole deformation, whereas enhanced pairing strength reduces $\beta_3$ and $\Delta E_{\mathrm{oct}}$, making the PES softer against octupole deformation. Within a certain range of pairing strength, however, $\beta_3$ exhibits little variation, while $\Delta E_{\mathrm{oct}}$ changes considerably. This indicates that the energy gain of octupole defomation $\Delta E_{\mathrm{oct}}$ is more sensitive to variations in the pairing strength than the ground-state deformation $\beta_3$. When the pairing strength exceeds a critical threshold, $\beta_3$ drops sharply to zero, signifying a transition from an octupole-deformed to a non-octupole-deformed ground state. For instance, when the neutron pairing strength $G_n$ is increased to $337\ \mathrm{MeV\cdot fm^3}$, $\beta_3$ vanishes abruptly for both $^{222}\mathrm{Ra}$ and $^{224}\mathrm{Ra}$, indicating the complete disappearance of octupole deformation. These results demonstrate that pairing correlations play an important role in shaping the octupole deformation of Ra isotopes.

\begin{figure}[htbp]
\includegraphics[width=7.6cm]{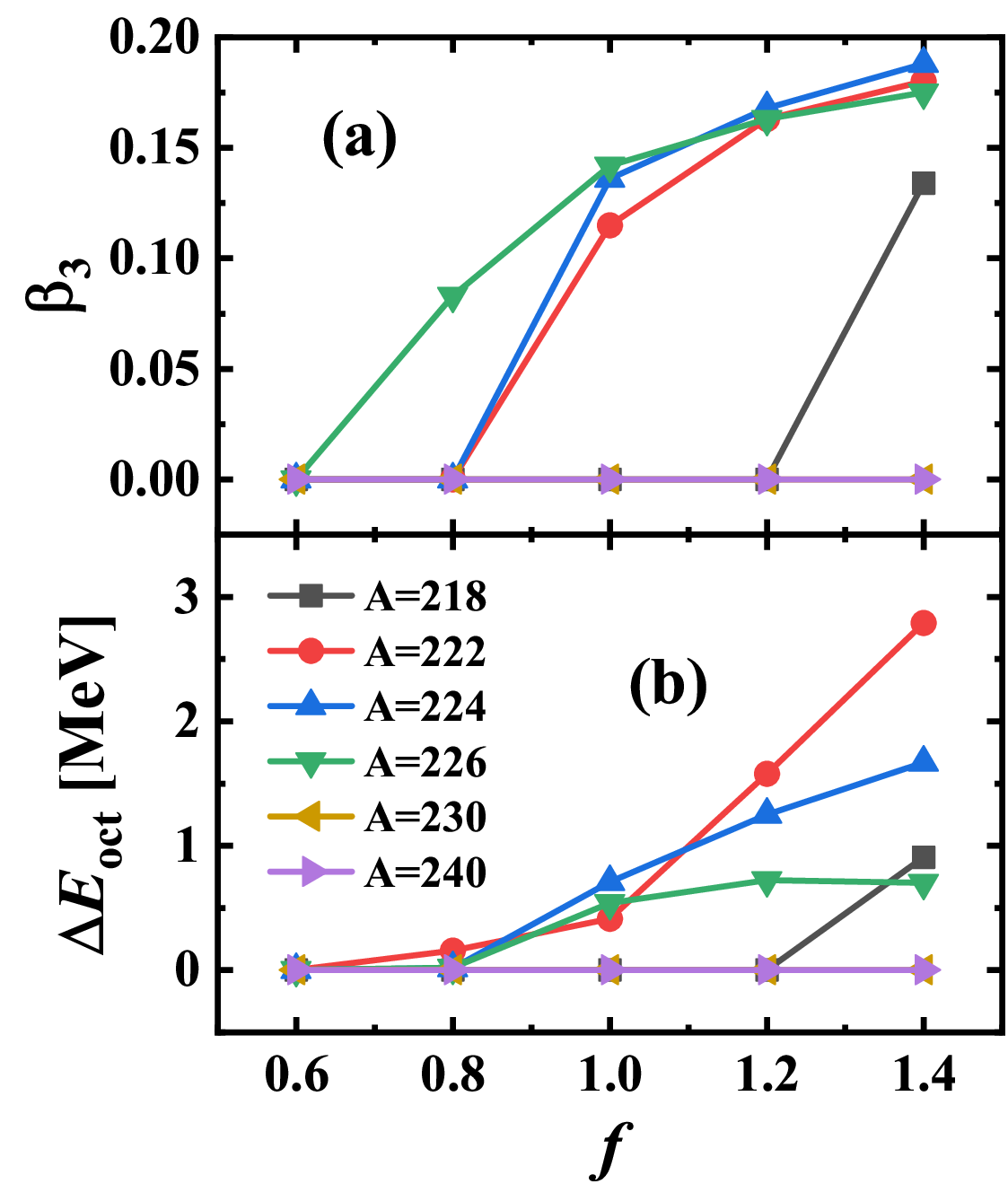}
\caption{(Color online) The calculated ground-state octupole deformation $\beta_3$ and the octupole energy gain $\Delta E_{\mathrm{oct}}$ as functions of the scaling factor $f$ of tensor coupling strength for Ra isotopes. Results are obtained from CDFT calculations in 3D lattice space using the PCF-PK1 functional.}
\label{Fig:f}
\end{figure}
The tensor coupling improves the description of the single-particle level and therefore can play an important role in the formation of octupole deformation. Figure~\ref{Fig:f} displays $\beta_3$ and $\Delta E_{\mathrm{oct}}$ as functions of the scaling factor $f$ of tensor coupling strength. For nuclei that are octupole deformed at the original coupling strength ($f = 1$), such as $^{222,224,226}\mathrm{Ra}$, increasing $f$ enhances both $\beta_3$ and $\Delta E_{\mathrm{oct}}$, indicating a more stable octupole deformation. In contrast, reducing the tensor coupling strength drives these nuclei towards a non-octupole-deformed shape. Notably, for nuclei that are not octupole deformed at $f = 1$, such as $^{218}\mathrm{Ra}$, increasing $f$ to 1.4 transforms the ground-state shape from a spherical shape to an octupole deformation.

\begin{figure}[htbp]
\includegraphics[width=6.6cm]{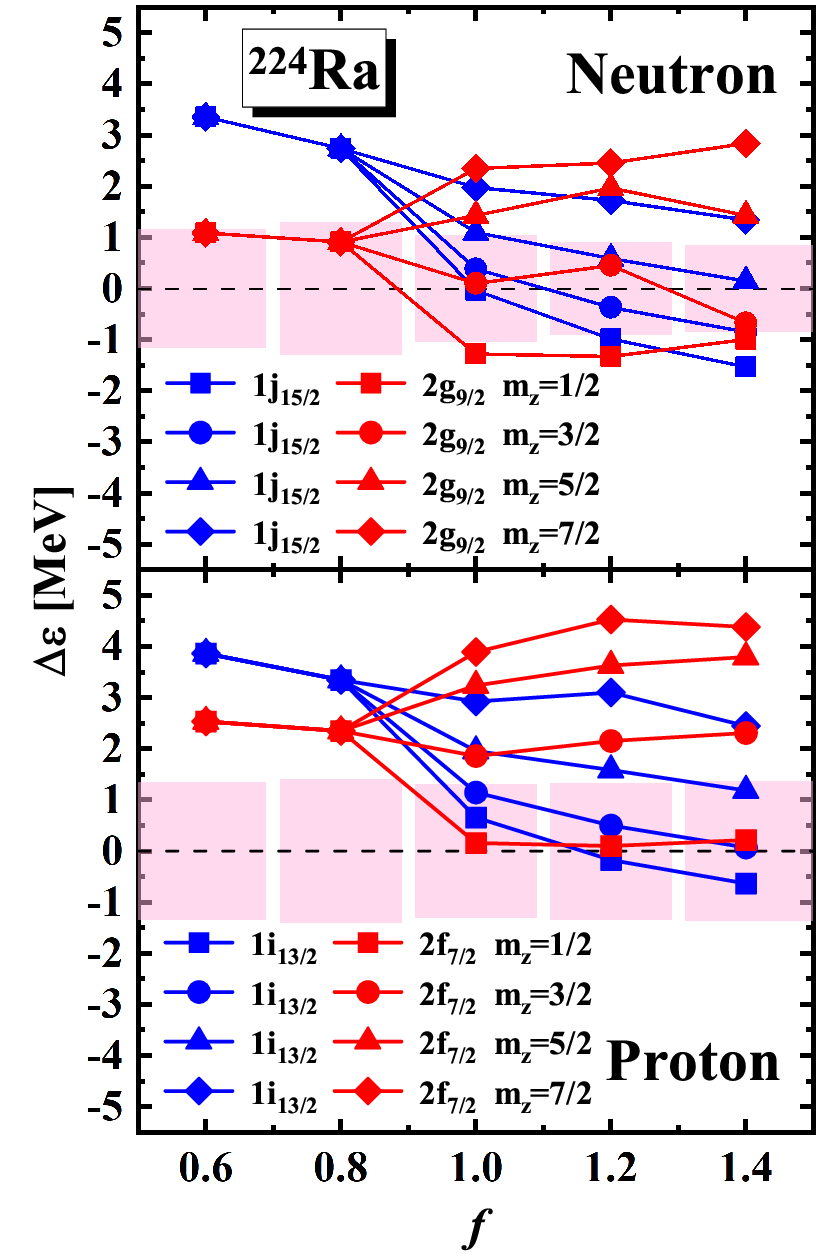}
\caption{(Color online) Single-particle energies of neutron orbitals from $\nu 1j_{15/2}$ and $\nu 2g_{9/2}$, and proton orbitals from $\pi 1i_{13/2}$ and $\pi 2f_{7/2}$, with $m_z = 1/2$, $3/2$, $5/2$ and $7/2$, near the Fermi surface for $^{224}$Ra, as functions of the scaling factor $f$ of tensor coupling strength, relative to the Fermi level. The results are obtained from covariant density functional theory calculations in three-dimensional lattice space using the PCF-PK1 functional, with the octupole deformation constrained to $\beta_3 = 0$ and the quadrupole deformation $\beta_2$ optimized to the energy minimum. The pink shaded band indicates the corresponding pairing gaps.}
\label{Fig:fRa224}
\end{figure}
To understand the microscopic mechanism of the tensor coupling effect on the formation of octupole deformation, we study the evolution of the single-particle energies and pairing gaps of $^{224}\mathrm{Ra}$ as functions of $f$. As shown in Fig.~\ref{Fig:fRa224}, the neutron pairing gap generally decreases with increasing $f$, indicating that enhancing the tensor coupling strength weakens the neutron pairing correlations, which favors the formation of octupole deformation. In contrast, the proton pairing gap exhibits little dependence on $f$. As $f$ increases, the high-angular-momentum intruder orbitals $\nu 1j_{15/2}$ and $\pi 1i_{13/2}$ decrease steadily, while the $m_z$ orbitals from $\nu 2g_{9/2}$ and $\pi 2f_{7/2}$ exhibit distinct behaviors: some $m_z$ orbitals increase and others decrease. Therefore, the energies and occupation probabilities of these octupole-driving orbitals undergo systematic changes near the Fermi surface as $f$ increases, allowing different $m_z$ orbitals to participate more effectively in the octupole coupling. This may account for the increase in both $\beta_3$ and $\Delta E_{\mathrm{oct}}$ with increasing tensor coupling strength.

\section{Summary and perspectives}
In summary, nuclear shape evolution in even-even Ra isotopes is investigated using CDFT with localized exchange terms in a 3D lattice space. Well-developed axial octupole deformations are found in $^{222-228}$Ra with no evidence of triaxial shapes. The energy gain of octupole deformation is employed to assess the stability of octupole deformation, with relatively larger values observed for $^{224}$Ra and $^{226}$Ra. A simplified analysis method based on the single-particle spectrum at the octupole deformation parameter $\beta_3 = 0$ is proposed to identify the key single-particle levels driving octupole deformation. It is found that the $m_z = 3/2$ orbitals from $\nu 1j_{15/2}$ and $\nu 2g_{9/2}$ and the $m_z = 1/2$ orbitals from $\pi 1i_{13/2}$ and $\pi 2f_{7/2}$ play crucial roles in the formation of octupole deformation in Ra isotopes. In $^{222-228}$Ra, these orbitals are located close to the Fermi surface with a small energy difference, thereby providing favorable conditions for octupole coupling. The pairing correlations and tensor coupling are found to have pronounced effects on octupole deformation. Weakening the pairing strength increases both $\beta_3$ and $\Delta E_{\mathrm{oct}}$, while strengthening the pairing reduces them. Enhancing the tensor coupling strength increases both $\beta_3$ and $\Delta E_{\mathrm{oct}}$ for nuclei already octupole deformed at the original strength, and can even induce octupole deformation for $^{218}$Ra that is not octupole deformed at the original strength. The increase in both $\beta_3$ and $\Delta E_{\mathrm{oct}}$ with stronger tensor coupling can be attributed to the systematic changes in the energies and occupation probabilities of the octupole-driving orbitals near the Fermi surface, which allow different $m_z$ orbitals to participate more effectively in octupole coupling. In the future, the 3D lattice CDFT with the PCF-PK1 functional can be extended to study octupole deformation in other isotopic chains and to further explore the coexistence and competition between octupole deformation and other shape degrees of freedom, such as tetrahedral and triaxial shapes. These results can provide rich and reliable nuclear properties for our studies of some exotic nuclear structure phenomena.

\section*{Acknowledgements}
This work was partly supported by the National Natural Science Foundation of China under Grants No. 12375109 and No. 11875070, the Fundamental Research Funds for the Central Universities under Grant No. 010-63263117, the Anhui project (Z010118169), the Key Research Foundation of Education Ministry of Anhui Province (2023AH050095), and the funding of the China Institute of Atomic Energy under Grants No. YC010270525794 and No. PA010271225779.



\end{CJK*}
\end{document}